\documentclass[seceq]{ptptex}
\usepackage{graphicx}
\newcommand{\mbf}[1]{\boldsymbol{#1}}

\title{Neutrino Oscillations in Gravitational Fields}

\author{
Hisae {\sc Maiwa} and
Shigefumi {\sc Naka}
}

\inst{
Department of Physics, College of Science and Technology \\
Nihon University, Tokyo 101-8308,Japan
}

\abst{ We study the quantum mechanical phases associated with the propagation of neutrinos in gravitational fields under the WKB approximation for Dirac equation. Applying the result, the investigations are made for gravitational effects on the neutrino oscillations, the lensing of neutrinos and so on.
}

\begin{document}

\maketitle

\section{Introduction}

The neutrino oscillation seems, nowadays, to be established experimentally as a fact in nature. The physics associated with the neutrino oscillation is still interesting subject to study from both theoretical and experimental points of view. Among those, the gravitational effects on the oscillation are, sometimes, considered to lead insignificant corrections, since those are very weak for the present status of neutrino observation. Nevertheless, such an effect inspires several physical interests to us.

For example, the breakdown of the equivalence principle\cite{Violation} was often discussed from the viewpoints of the influence on the oscillation as a matter like effect and of a basic problem in gravity. In those approaches, the breakdown is usually introduced by assuming tiny mass dependence of gravitational coupling constant, though the way of the breakdown is not unique and may not always be natural, since the breakdown is usually set after the standard geometrical approach to gravity that holds under the equivalence principle. 

Another interesting gravitational effect on neutrinos may be the gravitational lensing or the gravitational redshift, which can be observed as real effects for photons\cite{Lensing}; in this case, the equivalence principle must be assumed to derive those effects. One can expect to observe a similar gravitational effect in neutrinos too\cite{Neutrino-lensing}. Namely, neutrinos will serve as an observational tool in astrophysics; and conversely, astrophysics will give an insight into neutrino physics. For example, if it is possible to observe simultaneously the deflection of the neutrinos and that of photons started from the same astrophysical object, then the masses of neutrinos will be determined instead of those differences. 

To study the neutrino oscillations in gravitational fields\cite{Oscillations-in-CST,Oscillations-in-CST-2}, we have to somehow evaluate the Dirac wave functions in curved spacetime. As a useful way to do this, the form \cite{heuristic-way} $\frac{1}{\hbar}\int_A^Bdx^\mu p_\mu(x)$ is sometimes used as the heuristic gravitationally induced wave function phase without saying on its validity. Here, $p_\mu(x)$ is the four-momentum of neutrino at $x$. The wave function with such a heuristic phase is obtained in an approximate solution of the Dirac equation in truth. The purpose of this paper is, thus, to formulate the propagation of neutrino in gravitational fields by using an approximation like the WKB method for the Dirac equation, which was discussed by Pauli\cite{Pauli} to study the electrons in external electromagnetic fields, with the application to solvable examples. The heuristic gravity induced phase becomes to have clear meaning in this approach. In the present paper, the equivalence principle is also assumed to hold. 

 The formulation will be done in the next section within the framework of two generations for sake of simplicity. In section 3, we apply the resultant wave functions to neutrinos in gravitational fields associated with the Schwarzschild exterior and interior solutions. Section 4 is devoted to discussions: the gravitational correction for oscillation phases, deflection angles of neutrinos and its relation to that of photons, and other discussion on practical problems.

In appendix A, we give an introduction to the approximation of the Dirac equation in a curved spacetime according to Pauli's method. Some calculations associated with the interior Schwarzschild geometry are also given in appendix B.

\section{The phases of neutrinos propagating curved spacetime} 

The neutrinos propagate vacuum oscillating among their flavor eigenstates $|\nu_a\rangle$, $(a=e,\mu,\tau)$, when those are superposition of mass eigenstates $|\nu_i\rangle,~(i=1,2,3)$. Writing the states in momentum representation by $|\tilde{\nu}\rangle$, the flavor eigenstates can be expressed as the superposition $|\tilde{\nu}_a\rangle=\sum_i U_{ai}|\tilde{\nu}_i\rangle$ using a global unitary matrix $[U_{ai}]$. Here, $|\tilde{\nu}_i\rangle$ is the state satisfying $(p_i\hspace{-0.8em}/-m_i)|\tilde{\nu}_i\rangle=0$ and $\langle \bar{\tilde{\nu}}_i|\tilde{\nu}_j\rangle =\delta_{ij}$ for $p_i=(E,\mbf{p}_i),~(\mbf{p}_i^Oscillat=E^2-m_i^2)$
\footnote{ We use the unit $\hbar=1$ except Appendix A. }. 
Then, the evolution of those states $|\tilde{\nu}_a\rangle$ from $x_A$ to $x_B$ in vacuum give rise to the phases so that
\begin{equation}
 |\nu_a(B)\rangle=\sum_iU_{ai}e^{-ip_i\cdot(x_B-x_A)}|\tilde{\nu}_i\rangle \label{evolution-1}
\end{equation}
In what follows, for simplicity, we confine the discussion to the framework of two generations; and so, the unitary mixing matrix has the form
\begin{equation}
 [U_{ai}]=\begin{bmatrix}\cos\theta & \sin\theta \cr -\sin\theta & \cos\theta
          \end{bmatrix}
\end{equation}
The phases in Eq.(\ref{evolution-1}) acquire another forms when neutrinos pass through matter field or a curved spacetime with a metric $g_{\mu\nu}\neq \eta_{\mu\nu}$. In the latter case, the Dirac equations for the mass eigenstates of neutrinos become
\begin{equation}
 [i\gamma^\alpha e^\mu_{(\alpha)} D_\mu -m_ic]|\nu_i\rangle = 0 , \label{Dirac/curved}
\end{equation}
where $e^\mu_{(\alpha)}$ and $D_\mu \equiv \partial_\mu-\frac{i}{2}\sigma^{\alpha\beta}\omega_{(\alpha)(\beta)\mu}$ are inverse vierbein and the covariant derivative for local Lorentz transformations, respectively. 

Hereafter, we will discuss the case that the spin connections $\omega_{(\alpha)(\beta)\mu}$ are negligible compared with the derivative term $\partial_\mu$
\footnote{
The ratio of $\omega$ to $\partial$ is the order of $\frac{r_g}{R^2}\frac{\hbar c}{E}\ll \frac{\hbar c}{ER}$, where $r_g$ is the gravitational scale, such as the Schwarzschild radius, of an astrophysical object, to which neutrinos pass around with distance $R$. Here, $E$ is the energy of neutrinos; and for example, the ratio becomes $\ll 10^{-27}$ for $E\sim 10Gev$ with $R\sim 10^5 km$, the sun like radius.
}. 
Then, the $|\nu_i\rangle$ in Eq.(\ref{Dirac/curved}) can be solved at $x_B$ in the form $|\nu_i(B)\rangle \simeq e^{iS(B,A)}|\nu_i^0(B)\rangle$ in the sense of WKB approximation (Appendix A). Here, $S_i(B,A)$ is a solution of the Hamilton-Jacobi equation
\begin{equation}
 g^{\mu\nu}(\partial_\mu S_i)(\partial_\nu S_i)-(m_ic)^2=0   \label{H-J-eq}
\end{equation}
at $x_B$; and $|\nu_i^0(B)\rangle$ is a state satisfying
\begin{equation}
 (\gamma^\alpha e_{(\alpha)}^\mu \partial_\mu S_i + m_ic)|\nu_i^0(B)\rangle=0 
\end{equation}
at the same spacetime point $x_B$. We, here, require $S_i(B,A) \rightarrow 0,~(x_B \rightarrow x_A)$ so that $S_i(B,A)$ tends to $-p_i\cdot(x_b-x_A)$ in flat spacetime limit. In other words, $|\nu_i(B)\rangle$ will tend to $e^{-ip_i\cdot(x_B-x_A)}|\tilde{\nu}_i\rangle$ in that limit. Thus, in the curved spacetime, we can write the counterpart of Eq.(\ref{evolution-1}) as
\begin{equation}
|\nu_a(B,A)\rangle \simeq \sum_iU_{ai}e^{iS_i(B,A)}|\nu_i^0(B)\rangle . \label{evolution-2}
\end{equation}
The equality $\simeq$ means that $|\nu_a(B,A)\rangle$ is an approximate solution of the Dirac equation at $x_B$ disregarding $\hbar\partial_\mu|\nu_i^0(B)\rangle$ as a negligible quantity compared with $\partial_\mu S_i(B,A)$. It should be noticed, however, that, if $x_B$ is a point in the asymptotic region such as $e_{(\alpha)}^\mu=\delta_\alpha^\mu$ and $\partial_\mu S_i=const.$, we may identify $|\nu_i^0(B)\rangle$ with $|\tilde{\nu}_i(B)\rangle$, although $S_i$'s are functions of $x_A$ and $x_B$. In such a case, Eq.(\ref{evolution-2}) is a right solution of the Dirac equation in the curved spacetime.

Hereafter, we assume that the $x_B$ is an observational point of neutrinos located at flat spacetime region and apply to $\{|\nu_i^0(B)\rangle\}$ the same normalization as $\{|\tilde{\nu}_i(B)\rangle\}$. Therefore, we obtain the following expression for the transition probability in the curved spacetime:
\begin{eqnarray}
 P(a \rightarrow b) &=& |\langle \tilde{\nu}_b(B)|\nu_a(B,A)\rangle|^2 \nonumber \\
 &=& \sum_{i,j}(U_{ai}U^*_{aj})(U^*_{bi}U_{bj})e^{i(S_i(B,A)-S_j(B,A))}
  \label{Pr A to B}
\end{eqnarray}
In particular, we have
\begin{equation}
 P(\nu_e \rightarrow \nu_e)=1-\sin^2(2\theta)\sin^2\left(\frac{\Delta S}{2}\right), ~(\Delta S=S_1-S_2) \label{Pr e to e}
\end{equation}
We, further, confine the discussion to the neutrinos go through a static gravitational field on the way. Then the energy of neutrinos conserve; that is, the time variable in the phase function $S_i$ can be separated at any spacetime point $x=(t,\mbf{x})$ as
\begin{equation}
 S_i(x)=-Et+\bar{S}_i(\mbf{x}), \label{eikonal-1}
\end{equation}
In this case, the $\Delta S=S_1-S_2$ in Eq.(\ref{Pr e to e}) is replaced by $\Delta\bar{S}=\bar{S}_1-\bar{S}_2$, and the transition amplitude becomes time independent. 

\section{Gravitational effects in neutrino oscillations}

We, here, discuss the phase $S_i$ of the mass $m_i$ neutrino in Eq.(\ref{evolution-2}) due to the static spherically symmetric gravitational field produced by a spherically symmetric mass $M$ body of radius $R$. We assume that the body is at rest against both the source of neutrinos at $A$ and observer at $B$. Then we may take the body as the origin of spatial coordinates, the spherical coordinates $(r,\theta, \phi)$, in which the $A$ and $B$ are located at places with the radial coordinates $R_A$ and $R_B$, respectively.

To solve the $S_i$, we have to study separately the equations for $S_i$, which hold outside and inside the body producing gravitational field. \\

i) $S_i$ for the exterior Schwarzschild solution. \\

The Schwarzschild metric, which holds outside the spherically symmetric body at rest, is defined by the diagonal vierbein\cite{Landau-Lifshitz}\cite{Misner-Thorne-Wheeler}

\begin{equation}
e^{(t)}_t=\left(1-\frac{r_g}{r}\right)^{\frac{1}{2}},~e^{(r)}_r=\left(1-\frac{r_g}{r}\right)^{-\frac{1}{2}},~e^{(\phi)}_\phi=r\sin\vartheta,~e^{(\vartheta)}_\theta=r,
\end{equation} \\
where $x^1=r,x^2=\vartheta,x^3=\phi$ are spherical coordinates, and $r_g=2GM/c^2$ is the Schwarzschild radius of the body. Since, the angular momentum $L$ of neutrinos conserve in the spacetime under consideration, we may choose $z$ axis as the direction of the angular momentum. In other words, the neutrinos are on the plane with $\vartheta=\frac{\pi}{2}$. Then, we can rewrite the phase function $S$ in Eq.(\ref{eikonal-1}) as
\begin{eqnarray}
S_i=-Et+L_i\phi+S_{ir}(r).  \label{eikonal-2}
\end{eqnarray}
We note that each neutrino $\nu_i$ may have different angular momentum $L_i$, since the momenta of respective particles $p_i=\sqrt{(E/c)^2-(m_ic)^2},(i=1,2)$ at $x_A$ are different for $m_1\neq m_2$. Substituting the expression (\ref{eikonal-2}) for Eq.(\ref{H-J-eq}), we have
\begin{eqnarray}
 \left(1-\frac{r_g}{r}\right)^{-1}\frac{E^2}{c^2} -\left(1-\frac{r_g}{r}\right)\left(\frac{dS_{ir}}{dr}\right)^2 -\frac{L_i^2}{r^2}-(m_ic)^2 = 0. 
  \label{exterior eq}
\end{eqnarray}
 This equation can be integrated to give
\begin{eqnarray}
 S_{ir} &=& \pm \int_A^B dr \sqrt{ \left(1-\frac{r_g}{r} \right)^{-2}\frac{E^2}{c^2}-\left(1-\frac{r_g}{r} \right)^{-1}\left(\frac{L_i^2}{r^2}+m_i^2c^2 \right)} \nonumber \\
 &\simeq& \pm \int_A^B dr \left[ \sqrt{p_i^2-\frac{L_i^2}{r^2}}+\frac{r_g}{2r}\left(p_i^2+\frac{E^2}{c^2}-\frac{L_i^2}{r^2}\right) \left/ \sqrt{p_i^2-\frac{L_i^2}{r^2}} \right. \right],  \label{exterior S_ri}
\end{eqnarray}
under the approximation up to the order of $\frac{r_g}{r}$. Here, $+$ and $-$ represent the integral along a $r$ increasing path and a $r$ decreasing path, respectively. From this expression for $S_{ir}$, the lower bound of $r$ integration becomes $\rho_i=L_i/p_i$, which should be read as the impact parameter of neutrinos determined at $x_A$; and so, we can put $\rho_1=\rho_2=\rho_0$. Then writing $\tilde{r}=r/\rho_0,\tilde{r}_g=r_g/\rho_0$ and so forth, the right-hand side of Eq.(\ref{exterior S_ri}) becomes

\begin{figure}
 \centerline{\includegraphics[width=6.8cm,height=4cm]{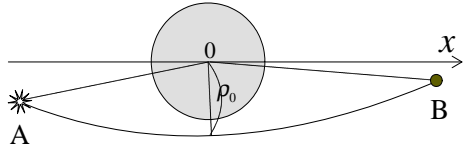}}
  \caption{The path from A(source) to B(observer).}
 \label{fig:1}
\end{figure}

\begin{equation}
 (r.h.s)=\sum_{k=A,B} L_i\int_1^{\tilde{R}_k} d\tilde{r}\left[ \sqrt{1-\frac{1}{\tilde{r}^2}}+\frac{\tilde{r}_g}{2}\frac{1+(\frac{E}{p_ic})^2}{\sqrt{\tilde{r}^2-1}}-\frac{\tilde{r}_g}{2}\frac{d}{d\tilde{r}}\sqrt{1-\frac{1}{\tilde{r}^2}} ~\right] \label{integral_1}
\end{equation}
under the condition $R_{A,B} \gg \rho_0 \gg r_g$. Here the $R_A$ and $R_B$ are radial coordinates of neutrinos at initial and final points, respectively. The integral of each term in Eq.(\ref{integral_1}) can be simply carried out, and we have
\begin{eqnarray}
 S_{ir}(B,A) & \simeq & \sum_{k=A,B}L_i\left[\left\{\sqrt{\tilde{R}_k^2-1}-\cos^{-1}\left(\frac{1}{\tilde{R}_k}\right)\right\} \right. \nonumber \\
 &+& \left. \frac{\tilde{r}_g}{2}\left\{1+\left(\frac{E}{p_ic}\right)^2\right\}\cosh^{-1}(\tilde{R}_k) -\frac{\tilde{r}_g}{2}\sqrt{1-\frac{1}{\tilde{R}_k^2}} ~\right] . \label{integral_2}
\end{eqnarray}

Now, $\phi_k^{(0)} = \cos^{-1}(\frac{1}{\tilde{R}_k}),(k=A,B)$ and $l_{AB}=\frac{L_i}{p_i}\sum_{k=A,B}\sqrt{\tilde{R}_k^2-1}$ are respectively $\phi$ coordinates of $A,B$ and the distance between $A,B$ in the flat spacetime. In a curved spacetime, the $\phi$ coordinate of neutrino $\nu_i$ as the function of $r$ is defined by\cite{Landau-Lifshitz}\cite{Mechanics-Landau-Lifshitz} $\frac{\partial S_i}{\partial L_i}=\phi_i+\frac{\partial S_{ir}}{\partial L_i}=0$. Then $\phi_{i,AB}=-\frac{\partial S_{ri}(B,A)}{\partial L_i}$ becomes the difference $\phi_b-\phi_A$ for the mass $m_i$ neutrino, to which a little calculation leads to
\begin{eqnarray}
 \phi_{i,AB} &=& \pm \int dr\left[ \frac{L_i}{r^2}\frac{1}{\sqrt{p_i^2-\frac{L_i^2}{r^2}}} - \frac{r_g}{2r}\left\{-\frac{2L_i}{r^2}\frac{1}{\sqrt{p_i^2-\frac{L_i^2}{r^2}}} + \frac{\left(p_i^2+\frac{E^2}{c^2}-\frac{L_i^2}{r^2}\right)\frac{L_i}{r^2}}{\left(p_i^2-\frac{L_i^2}{r^2}\right)^{\frac{3}{2}} } \right\} \right] \nonumber \\
 &=& \phi^{(0)}_{AB}+ \frac{\tilde{r}_g}{2}\sum_{k=A,B}\frac{1+\left(\frac{E}{cp_i}\right)^2-\frac{1}{\tilde{R}_k^2}}{\sqrt{1-\frac{1}{\tilde{R}_k^2}}}. \label{deflection}
\end{eqnarray}
Here the second term in the right-hand side of Eq.(\ref{deflection}) is nothing but the deflection angle $\delta\phi_{i,AB}$ of the mass $m_i$ neutrino due to the gravitation. In practical problems, it will be sufficient to make approximate $1+\left(\frac{E}{p_ic}\right)^2\simeq 2+\left(\frac{m_ic^2}{E}\right)^2,~\frac{1}{\tilde{R}_k^2}\simeq 0$ and $\cosh^{-1}(\tilde{R}_k) \simeq \log(2\tilde{R}_k)$ in equations (\ref{integral_2}) and (\ref{deflection}). Under these approximations, we can write the deflection angle as $\delta\phi_{i,AB}=2\tilde{r}_g(1+\frac{m_i^2c^4}{2E^2})$
\footnote{
The deflection angle $\delta\phi_{i,AB}$ will tend to that of photon $\delta\phi_{AB}^{ph}=2\tilde{r}_g$ for $m_i\rightarrow 0, \tilde{R}_k\rightarrow \infty$
}.

Substituting Eq.(\ref{deflection}) for Eq.(\ref{eikonal-1}), the phase function in Eq.(\ref{eikonal-2}) can be evaluated as
\begin{equation}
 L_i\phi_{AB}+S_{ir}(B,A)=p_il_{AB}+L_i\delta\phi_{i,AB}+L_i\left\{\frac{1}{2}\delta\phi_{i,AB}\log\left(4\tilde{R}_A\tilde{R}_B\right)-\tilde{r}_g \right\}. 
\label{S bar}
\end{equation}
Here, taking $L_i\tilde{r}_g\simeq \left(\frac{E}{c}-\frac{m_i^2c^3}{2E}\right)r_g$ and $L_i\delta\phi_{i,AB}\simeq 2r_g\left(\frac{E}{c}\right)$ into account, we finally derive the phase difference
\footnote{ If we may put $\phi_{AB}=\sum_{k=A,B}\cos^{-1}(\frac{1}{\tilde{R}_k})$, then $\Delta S$ can be calculated directly from Eq.(\ref{integral_2}). The $\Delta S$ in this case is simply obtained by the substitution $r_g \rightarrow r_g\{1-\log(4\tilde{R}_A\tilde{R}_B)\}$ in Eq.(\ref{Delta S out}).}
\begin{equation}
 \Delta S \simeq \frac{\Delta m^2 c^3}{2E}(l_{A,B} - r_g), \label{Delta S out}
\end{equation}
within the approximation up to the order $\Delta m^2=m_2^2-m_1^2$. It may be surprising that the $\Delta S$ looked as if it does not depend on $R_A,R_B$ but on the Schwarzschild radius $r_g$ of the object, around which the neutrinos pass through. In reality, however, these parameters are included in the definition of $l_{AB}=\sum_{k=A,B}\sqrt{R_k^2-\rho_0^2}$. \\

ii) $S_i$ for the interior Schwarzschild solution. \\

For the neutrino passing through an astrophysical object, we can suppose two types of situations. One is the case such that neutrinos pass across the object on a long way from $A$ to $B$. The other is the long-baseline accelerator experiment on the Earth. In both cases, we have to consider the matter effects in addition to gravitational effects. Then, the phase difference $\Delta S$ calculated from gravitational effect only should be understood as a supplementary result. \\

\begin{figure}
 \centerline{\includegraphics[width=5cm,height=2.8cm]{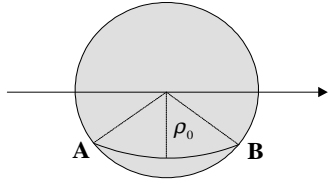}}
  \caption{The path inside a homogeneous sphere.}
 \label{fig:2}
\end{figure}

As for the second case, however, it may be worthwhile to show the result obtained under the condition that the Earth is a homogeneous sphere of incompressible fluid with radius $R$. The vierbeins are, then, given by

\begin{equation}
 e^{(t)}_t = \left(\frac{3}{2}\sqrt{1-\frac{r_g}{R}}-\frac{1}{2}\sqrt{1-\frac{r_gr^2}{R^3}}\right), ~e^{(r)}_r = \left(1-\frac{r_gr^2}{R^3}\right)^{-\frac{1}{2}}.
\end{equation}
The $e^{(\phi)}_\phi=\sin\theta$ and $e^{(\theta)}_\theta=r$ are common to the exterior Schwarzschild solution. Let us consider the situation of Fig.2 for the trajectories of neutrinos; that is, $R_A=R_B=R\simeq \rho_0 \gg r_g$. Then the following can be obtained (Appendix B):

\begin{equation}
 \Delta S \simeq \frac{\Delta m^2c^3}{2E}l_{AB}\left(1-\frac{r_g}{2\rho_0}\right). \label{Delta S in}
\end{equation}
In this case, we have to read Eqs.(\ref{evolution-2}) and (\ref{Pr e to e}) as those of semiclassical approximation, since we deal with that the gravity remains on the surface of the Earth.

\section{Summary and Discussion}

In this paper, we have studied mainly two problems of neutrino oscillation: the formulation of neutrino wave function in gravitational fields, its application to simple examples of gravitational fields. First, to formulate the wave function in gravitational fields, we followed the semiclassical method for the Dirac equation, which was discussed by Pauli in a problem of electrons under electromagnetic fields. The resultant Dirac wave function
is obtained from that of the flat spacetime by substituting the phase for Hamilton's principal function of the neutrino in the curved spaceitme, providing the observation point is located at the flat spacetime. We can, thus, recognize that the heuristic phases used in curved spacetimes are nothing but the result of this approximation. 

Secondly, we attempted to apply our formula to the neutrinos with two generations for the gravitational fields in outside of a static uniform sphere and for one in inside of the sphere. The latter is applicable, for example, to such an experiment as the long-base line, although the effect is examined to be too small in practical problems. On the other side, in the former case, the gravitational effect yields the probability (\ref{Pr e to e}) characterized by the oscillation phase $\frac{\Delta S}{2}=\frac{\pi}{l_0}(l_{AB}-r_g)$ with the mixing angle $\theta$, where $l_0=\frac{4\pi E}{\Delta m^2 c^3}$ is the oscillation scale in the vacuum. The result says that if the neutrinos started from a source object such as a supernova passes through a gravitational area by a star with the scale of sun, then the effective distance will be shifted by $r_g\approx 3km$. This looks like that the effect of $r_g$ may be ignored compared with the distance $\sim10^5$ light-year, which the neutrinos travel across. We must be careful, however, on this problem, since $l_{AB}+r_g$ and its remainder divided by $l_0$ gives the same oscillation effect; and so, the long distance $l_{AB}$ play the role of well mixing of neutrinos between generations. Then, the $r_g$ dependence due to the pass dependence of neutrinos may give an interesting observational effect in the oscillation.

We note that the gravitational effect in the oscillation phase can be understood in terms of the gravitational lensing effect. Indeed, one can write the effect as $\frac{\pi}{l_0}r_g=\frac{1}{4}\rho_0\Delta\delta\phi$ with $\Delta\delta\phi=\delta\phi_1-\delta\phi_2$. Here, $\delta\phi_i,(i=1,2)$ are deflection angles for mass eigenstates of neutrinos, which relate to the deflection angle of the photon $\delta\phi^{ph}$ by $\delta\phi_i=\delta\phi^{ph}(1+\frac{m_i^2c^4}{2E^2})$. If we can observe both deflection angles of a massive neutrino and that of photon from the neutrino source object, one can determine $\Delta m_i^2$ in principle, although such an observation will be very hard in practice. We also note that we can introduce the dynamical time $t_i$, the time as a function of $r$, of each particle by $\frac{\partial S_i(B,A)}{\partial E}=0$ similar to the $\phi_i$ with deflection angle. The Eq.(\ref{S bar}) gives us the difference of these dynamical times as $\Delta t=t_{AB2}-t_{AB1}=\frac{\Delta m^2c^4}{2E^2}(\frac{l_{AB}}{c}-\frac{r_g}{c})$, in which the gravitational correction is again very small in practical problems. 

Throughout this paper, we discussed the neutrinos under the influence of gravitational fields only provided the mixing angles were constants; that is, that we did not take the matter effect into account. We also note that the oscillation phases derived in the Schwarzschild spacetime under the condition $\rho_0 \gg r_g$ can not be applied for neutrinos passing close to heavy astrophysical objects such as neutron stars, though Eqs.(\ref{eikonal-2}) and (\ref{exterior eq}) hold. In such a case\cite{Strong-gravity}, it becomes necessary to develop another approximation by taking the spin connections into account. Another basis of this paper is the equivalence principle, the universality of gravitational coupling. In our approximation, the gravitational effect appears only through $r_g$. Thus, the breakdown of the universality by $r_g \rightarrow r_{gi}$ simply leads to the substitution $\Delta(m^2r_g)$ for $\Delta m^2r_g$, for example, in Eq.(\ref{Delta S out}). Together with the extension of generations, there remain interesting future problems according to this line.

\section*{Acknowledgements}

The authors wish to express their thanks to the members of their laboratory for discussions and encouragement.

\appendix

\section{Spin $1/2$ particles in curved spacetime}

The curved spacetime can be described by the metric $g_{\mu\nu}=\eta_{\alpha\beta}e^{(\alpha)}_\mu e^{(\beta)}_\nu$ with the vierbein $\{ e^{(\alpha)}_\mu \},~(\alpha,\mu=0,1,2,3)$, where $\eta_{\alpha\beta}$ is the Minkowski metric. The vierbein is supposed to satisfy the parallelism condition

\begin{equation}
 \nabla_\mu e^{(\alpha)}_\nu \equiv \partial_\mu e^{(\alpha)}_\nu - \Gamma^\rho_{\nu\mu}e^{(\alpha)}_\rho + \omega^{(\alpha)}{}_{(\rho)\mu}e^{(\rho)}_\nu =0
\end{equation}
; that is, in terms of the inverse vierbein $e_{(\alpha)}^\mu e^{(\beta)}_\nu=\delta_\alpha^\beta$, 
\begin{equation}
 \nabla_\mu e_{(\alpha)}^\nu \equiv \partial_\mu e_{(\alpha)}^\nu + \Gamma^\nu_{\rho\mu}e_{(\alpha)}^\rho - \omega^{(\beta)}{}_{(\alpha)\mu}e_{(\beta)}^\nu =0 .
\end{equation}
Then, the Dirac equation in the curved spacetime has the form
\begin{eqnarray}
 [i\hbar\gamma^\alpha e^\mu_{(\alpha)} D_\mu -mc]\psi=0, \label{Dirac 1}
\end{eqnarray}
where $D_\mu \psi \equiv (\partial_\mu-\frac{i}{2}\sigma^{\alpha\beta}\omega_{(\alpha)(\beta)\mu})\psi,~(\sigma^{\alpha\beta}=\frac{i}{2}[\gamma^\alpha,\gamma^\beta])$ is the covariant derivative acting on the spinor $\psi$. Throughout this paper, we consider the case such that the spin connections $\{\omega_{(\alpha)(\beta)\mu} \}$ are negligible compared with the other terms in the Dirac equation, and so the $\psi$ satisfies
	\begin{eqnarray}
	[i\hbar\gamma^\alpha e^\mu_{(\alpha)} \partial_\mu -mc]\psi = 0. \label{Dirac 2}
	\end{eqnarray}

To obtain an approximate solution of this equation, let us apply the semi-classical method by Pauli\cite{Pauli}, the Dirac counterpart of the WKB approximation; and we put
	\begin{eqnarray}
	\psi=e^{\frac{iS(x)}{\hbar}}u,
	\end{eqnarray}
where $S$ and $u$ are a scalar phase function and a four components spinor, respectively. Substituting this expression for Eq.(\ref{Dirac 2}), we have
	\begin{eqnarray}
	[i\hbar\gamma^\alpha e^\mu_{(\alpha)}(\partial_\mu + \frac{i}{\hbar}\partial_\mu S) - mc]u = 0. \label{Dirac 3}
	\end{eqnarray}
In non-relativistic Schr\"odinger equation, the semiclassical approximation is carried out by expanding $S$ in powers of $\frac{\hbar}{i}$. On the other hand, in the Dirac equation, the $\frac{\hbar}{i}$ expansion is applied to $u$ instead of $S$\cite{Pauli}; that is,
\begin{equation}
 u=u_0+\left(\frac{\hbar}{i}\right)u_1+\left(\frac{\hbar}{i}\right)^2 u_2+ \cdots .
\end{equation}
Substituting the expansion to (\ref{Dirac 3}), we obtain the successive equations
\begin{eqnarray}
 & ( \gamma^\alpha e^\mu_{(\alpha)}\partial_\mu S + mc)u_0 = 0 , & \label{0-th} \\
 & ( \gamma^\alpha e^\mu_{(\alpha)}\partial_\mu S + mc)u_1=-\gamma^\alpha e^\mu_{(\alpha)}\partial_\mu u_0 \label{1-th}
\end{eqnarray}	
and so on, at each order of $\hbar$. Multiplying $(\gamma^\alpha e^\mu_{(\alpha)}\partial_\mu S - mc)$ from the left-hand side of Eq.(\ref{0-th}), we obtain the Hamilton-Jacobi equation
\begin{eqnarray}
	g^{\mu\nu}\partial_\mu S \partial_\nu S -m^2c^2=0.
	\end{eqnarray}
Using the $S$ and a 4-spinor $w$, the $u_0$ can be expressed as
\begin{eqnarray}
 u_0 & \propto & \Lambda_{-}(S)w , \label{projection} \\
 \Lambda_{-}(S) &=& \frac{-\gamma^\alpha e^\mu_{(\alpha)}\partial_\mu S + mc}{2mc}, 
\end{eqnarray}
where $\Lambda_{-}(S)$ is the projection operator for solutions of Eq.(\ref{0-th}). 

Now, under the lowest approximation
\begin{equation}
u\simeq u_0 e^{\frac{i}{\hbar}S} \label{0th}
\end{equation}
Eq.(\ref{1-th}) leads to $e_{(\alpha)}^\mu\partial_\mu(\bar{u}_0\gamma^\alpha u_0)=e_{(\alpha)}^\mu \partial_\mu(\bar{u}\gamma^\alpha u)=0$ by taking $\bar{u}_0(\gamma^\alpha e_{(\alpha)}^\mu\partial_\mu S+mc)=0$ into account. On the other hand, since we have disregarded the spin connection, the inverse vierbein satisfies $\nabla_\mu e_{(\alpha)}^\nu = \partial_\mu e_{(\alpha)}^\nu+\Gamma^\nu_{\mu\rho} e_{(\alpha)}^\rho=0$. Thus, we can write the continuity equation in the covariant form $e_{(\alpha)}^\mu \partial_\mu(\bar{u}\gamma^\alpha u)=\nabla_\mu(\bar{u}e_{(\alpha)}^\mu \gamma^\alpha u)=0$. This says that the above approximation (\ref{0th}) is appropriate for the spin $1/2$ particle in the gravitational field under consideration.

\section{ The phase function inside a homogeneous sphere.}

The phase functions $S_i$ have the form (\ref{eikonal-2}) in this case too; and the counterpart of Eq.(\ref{exterior eq}) becomes
\begin{equation}
 \left(\frac{3}{2}\sqrt{1-\frac{r_g}{R}}-\frac{1}{2}\sqrt{1-\frac{r_gr^2}{R^3}}\right)^{-2}\frac{E^2}{c^2}-\left(1-\frac{r_gr^2}{R^3}\right)\left(\frac{d S_{ri}}{dr}\right)^2-\frac{L_i^2}{r^2}-(m_ic)^2=0.
\end{equation}
Now, let us consider the case $r_g \ll \rho_0 \leq r \leq R$; then, within the approximation up to $\frac{r_g}{R}$, the $S_{ir}$ can be integrated to the following form:
\begin{eqnarray}
 S_{ir} &=& \pm \int \frac{dr}{\sqrt{1-\frac{r_gr^2}{R^3}}}\left\{ \left(\frac{3}{2}\sqrt{1-\frac{r_g}{R}}-\frac{1}{2}\sqrt{1-\frac{r_gr^2}{R^3}}\right)^{-2}\frac{E^2}{c^2}-\frac{L_i^2}{r^2}-m_i^2c^2 \right\}^{\frac{1}{2}} \nonumber \\
 &\simeq& \pm\int d\tilde{r} L_i\left\{ \sqrt{1-\frac{1}{\tilde{r}^2}}+\frac{\frac{3\tilde{r}_g}{2\tilde{R}}-\frac{\tilde{r}_g\tilde{r}^2}{2\tilde{R}^2}}{2\sqrt{1-\frac{1}{\tilde{r}^2}}}\left(\frac{E}{cp_i}\right)^2 + \frac{\tilde{r}_g\tilde{r}^2}{2\tilde{R}^3}\sqrt{1-\frac{1}{\tilde{r}^2}}   \right\} \nonumber \\
 &=& 2L_i\left[\left\{\sqrt{\tilde{R}^2-1}-\cos^{-1}\left(\frac{1}{\tilde{R}^2}\right)\right\} + \left(\frac{E}{cp_i}\right)^2\frac{\tilde{r}_g}{2\tilde{R}}\sqrt{\tilde{R}^2-1} \right. \nonumber \\
 &~& ~~~ + \left. \left\{1+\left(\frac{E}{cp_i}\right)^2\right\}\frac{\tilde{r}_g}{6\tilde{R}^3}(\tilde{R}^2-1)^{\frac{3}{2}} \right] \label{eikonal-3}
\end{eqnarray}

In the practical neutrino experiments connecting two points on the earth, the long baseline accelerator experiments, we may put $\tilde{R}\simeq 1$, since neutrinos pass through the ground near surface of the earth. Then the last term in the right-hand side of Eq.(\ref{eikonal-3}) can be negligible, and we obtain
\begin{eqnarray}
 S_{ir} &\simeq& 2L_i\left[\left\{\sqrt{\tilde{R}^2-1}-\cos^{-1}\left(\frac{1}{\tilde{R}}\right)\right\} + \left(\frac{E}{cp_i}\right)^2\frac{\tilde{r}_g}{2\tilde{R}}\sqrt{\tilde{R}^2-1} \right\} \\
 &=& p_il_{AB}-L_i\phi_{AB}+\frac{r_gl_{AB}}{R}\frac{1}{p_i}\left(\frac{E}{c}\right)^2,
\end{eqnarray}
where we have identified the geometrical angle $2\cos^{-1}\left(\frac{1}{\tilde{R}}\right)$ with the physical angle $\phi_{AB}(=-\frac{\partial S_{ir}}{\partial L_i})$, since the spacetime is almost flat on the earth. From this expression for $S_{ir},(i=1,2)$, we can derive
\begin{equation}
 \Delta S \simeq \frac{\Delta m^2c^3}{2E}l_{AB}\left(1-\frac{r_g}{2\rho_0}\right).
\end{equation}

%%%%%%%%%%%%%%%%%%%%%%%%%%%%%%%%%%%%%%%%%%%%%%%%%%%%%%%%%%%%%%%%%%%%%%%%%%%%%%%

\end{document}